\documentclass{elsart}
\usepackage{epsfig}
\def\be{\begin{eqnarray}}
\def\ee{\end{eqnarray}}
\newcommand{\mat}{\left ( \begin{array}{cc}}
\newcommand{\emat}{\end{array} \right )}
\newcommand{\vect}{\left ( \begin{array}{c}}
\newcommand{\evect}{\end{array} \right )}

\begin{document}
\begin{frontmatter}
\hyphenation{Coul-omb ei-gen-val-ue ei-gen-func-tion Ha-mil-to-ni-an
  trans-ver-sal mo-men-tum re-nor-ma-li-zed mas-ses sym-me-tri-za-tion
  dis-cre-ti-za-tion dia-go-na-li-za-tion in-ter-val pro-ba-bi-li-ty
  ha-dro-nic he-li-ci-ty Yu-ka-wa con-si-de-ra-tions spec-tra
  spec-trum cor-res-pond-ing-ly}
\title{Randomness on the Lattice}
%\footnote{Invited talk at Xth 
%International Light-Cone Meeting on Non-Perturbative QCD and Hadron
%Phenomenology, Heidelberg 12-17 June 2000. To appear in proceedings.}
%
\author{J.J.M. Verbaarschot}
\address{University at Stony Brook, NY 11794 \\
{\tt verbaarschot@nuclear.physics.sunysb.edu}}

\date{August 19 2000}
\begin{abstract}
In this lecture we review recent lattice QCD studies of the
statistical properties of the eigenvalues of the QCD Dirac 
operator. We find that the fluctuations of the smallest Dirac eigenvalues
are described by  chiral Random Matrix Theories with the global
symmetries of the QCD partition function.  Deviations from chiral
Random Matrix Theory beyond the Thouless energy can be 
understood analytically by means of partially quenched chiral
perturbation theory.
 
\end{abstract}
\end{frontmatter}

%%%%%%%%%%%%%%%%%%%%%%%%%%%%%%%%%%%%%%%%%%%%%%%%%%%%%%%%%%%%%%%%%%%%%%
%%%%%%%%%%%%%%%%%%%%%%%%%%%%%%%%%%%%%%%%%%%%%%%%%%%%%%%%%%%%%%%%%%%%%%
%%%%%%%%%%%%%%%%%%%%%%%%%%%%%%%%%%%%%%%%%%%%%%%%%%%%%%%%%%%%%%%%%%%%%%

\section{Introduction}
\label{intro}
A significant part of our understanding of nonperturbative phenomena in 
QCD, such as chiral symmetry breaking,  confinement or the existence
of nuclei, results from
simulations of the QCD partition function on a Euclidean space-time 
lattice (see \cite{DeTar,Creutz} for reviews). In spite of its numerous
successes, this approach has several disadvantages. One of them is
the use of Euclidean space-time which requires a highly nontrivial 
analytical continuation to Minkowski space-time. 
One of the promising approaches
that works  directly in a Hamiltonian framework 
is discrete light-cone QCD
\cite{dlcq}, but its results for 4-dimensional nonabelian
gauge theories can not yet compete with lattice QCD. 
A second disadvantage of lattice
QCD is that analytical understanding of most lattice data seems
beyond reach. Therefore, it is imperative to provide an analytical
explanation of lattice observables whenever possible. One such observable
is the Euclidean Dirac spectrum. We have
proved \cite{OTV,DOTV} our conjecture \cite{SVR,V} that the 
fluctuations of the smallest Dirac eigenvalues are given by a chiral
Random Matrix Theory (chRMT) 
with the global symmetries of the QCD partition function.
In this lecture we give a review of recent lattice simulations that
support this assertion. A recent comprehensive review of chiral Random
in QCD was given in \cite{Tilorev}.

Of course, chRMT cannot provide us with a $complete$ description of
the QCD Dirac spectrum. What is the domain
of validity of chRMT? To answer this question  we need to identify three
different scales in the Dirac spectrum. 
The first scale is the
smallest nonzero eigenvalue, $\lambda_{\rm min}$. Its average position 
is given by the mean level spacing, $\Delta(\lambda)$,  
which is the inverse of the
average spectral density near zero, $\rho(0)$,
\be
\lambda_{\rm min} \approx \Delta \lambda = \frac 1{\rho(0)} = \frac
\pi{\Sigma V}.
\ee
Because of the axial symmetry the nonzero eigenvalues of the Euclidean
Dirac operator $D$ appear in pairs $\pm \lambda_k$.
The average spectral density is then defined by 
$\rho(\lambda) = \langle \sum_k \delta(\lambda -\lambda_k) \rangle$, 
and $\pi \rho(0)/V$ (with $V$ the volume of space-time)
is identified as  the chiral condensate, $\Sigma$, through
the Banks-Casher formula \cite{BC}.
A second scale in the Dirac spectrum is the mass scale for which the
Compton wavelength of the corresponding Goldstone boson is equal to
the size of the box. This scale, also known as the Thouless
energy, is given by \cite{GL,LS,Vplb}
\be
m_c = \frac {F^2}{\Sigma L^2},
\label{thouless}
\ee
where $L$ is the linear size of the box. A third scale is the typical
hadronic mass scale given by $\Lambda_{\rm QCD}$.

Because QCD has a mass gap,  for volumes with $\Lambda_{\rm QCD} L \gg 1$,
the QCD partition function in the phase of spontaneous
broken chiral symmetry, 
 reduces to that of a gas of Goldstone bosons. 
For momenta and masses well below $\Lambda_{\rm QCD}$,
this effective chiral
partition function can be written down solely on the basis of
the global symmetries of QCD.  A further simplification arises for $m
\ll m_c$. In this domain the fluctuations of the constant fields are
much larger than the fluctuations of the nonzero momentum modes and
kinetic term of the chiral Lagrangian can be ignored in the
calculation of the mass dependence of the partition function
\cite{GL,LS}. 
This is
the domain of validity chiral Random Matrix Theory. A formal proof of
this statement \cite{Vplb,james,OTV,DOTV}
requires the introduction of additional ghost quarks
with spectral mass $z$ equal to the argument of the resolvent of the
Dirac operator. Because $z$ is a free parameter, it can always be chosen
 such that $z \ll m_c$, and the Dirac spectrum in this domain
is thus given by chiral Random Matrix Theory.

\section{Chiral Random Matrix Theory}

The chiral random matrix partition function
 with the global symmetries of the QCD partition function 
is defined by {\cite{SVR,V}}
\be
Z^\nu_\beta({M}) =
\int DW \prod_{f= 1}^{N_f} \det{\left (\begin{array}{cc} m_f & iW\\
iW^\dagger & m_f \end{array} \right )}
e^{-\frac{N \beta}4 \Sigma{\rm Tr}W^\dagger W },
\label{ranpart}
\ee
where $W$ is a $n\times m$ matrix with $\nu = |n-m|$ and
$N= n+m$.
As is the case in QCD, we assume that the equivalent of the topological charge
$\nu$ does not exceed $\sqrt N$,
so that, to a good approximation, $n = N/2$.
Then the parameter $\Sigma$ can be identified as the chiral condensate and
$N$ as the dimensionless volume of space time (Our units are defined
such that the density of the modes $N/V=1$). The chiral ensembles
are classified according to the Dyson index $\beta$.
The matrix elements of $W$ are either real ($\beta = 1$, chiral
Gaussian Orthogonal Ensemble (chGOE)), complex
($\beta = 2$, chiral Gaussian Unitary Ensemble (chGUE)),
or quaternion real ($\beta = 4$, chiral Gaussian Symplectic Ensemble (chGSE)).
For QCD with three or more colors and quarks in the fundamental representation
the matrix elements of the Dirac operator are complex and we have $\beta = 2$.
The ensembles with $\beta =1 $ and $\beta =4$ are relevant in the case
of two colors or adjoint fermions. For staggered fermions, the value of
the Dyson index in these two cases is reversed. 
The reason for choosing a Gaussian distribution of the matrix elements is
its mathematically simplicity. It can be shown that the correlations
of the eigenvalues on the scale of the average level spacing do not depend on
the details of the probability distribution 
\cite{Brez96,ADMN,Kanz97,senert,Guhr,Seneru,Kanzieper}.

\section{Lattice Results}
We start this section by stressing that only spectral 
properties on the scale of the
average level spacing can be described by Random Matrix
Theory. We thus unfold the spectrum by rescaling the eigenvalues according
to the macroscopic average level spacing obtained by averaging over
many consecutive levels inside a small but finite interval. 
Below we always discuss the statistical properties of the unfolded
eigenvalues, with average spectral density equal to unity. They have
been analyzed in several different ways.

\begin{figure}[!ht]
  \begin{center}
    \epsfig{figure=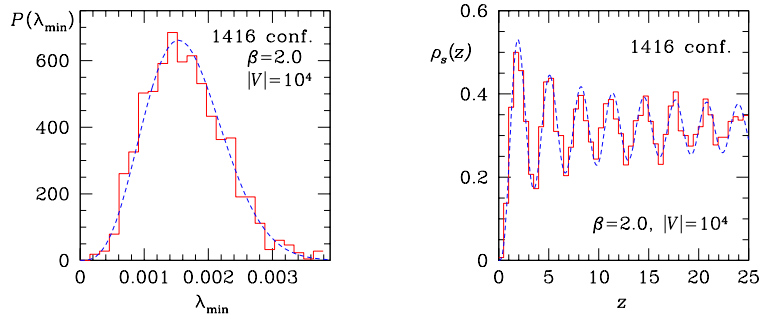,width=100mm}
    \caption{Distribution of the smallest eigenvalue (left) and
      microscopic spectral density (right) of the staggered Dirac
      operator in quenched SU(2).  The dashed curves are the predictions
      of the chSE for $N_f=0$ and $\nu=0$. (From Ref. \cite{Berbenni}.)}
    \label{figmicro}
  \end{center}
\end{figure}

First, by means of the microscopic spectral density defined by \cite{SVR}
\be
\rho_s(u) = \lim_{V\rightarrow \infty} \frac 1{V\Sigma} \langle
\rho(\frac u{V\Sigma})\rangle.
\label{rhosu}
\ee
For $\beta =2 $ it is given by \cite{VZ,Vinst}
\be
 \rho_s(z)  = \frac {z}{2} \left[J^2_{N_f+|\nu|}(z) -J_{N_f+|\nu|+1}(z)
  J_{N_f+|\nu|-1}(z)\right].
  \label{micro2}  
\ee
The result for $\beta =1$ \cite{Vnc2} and $\beta =4$ \cite{Forrester}
is more complicated but can be expressed as an integral over Bessel functions. 
The microscopic spectral density 
 was first observed for Dirac spectra of instanton liquid field
configurations \cite{Vinst} both for $N_c =2$ and $N_c = 3$. Its first
lattice studies, for quenched $SU(2)$ gauge theory with staggered
fermions (with Dyson index $\beta =1$), were performed in
\cite{Berbenni} (see  Fig. \ref{figmicro}). 
The agreement between lattice QCD at Random matrix
theory is equally good for $N_c =3$ \cite{poul3,tilo3} (with Dyson
index $\beta =2$), for
QCD with adjoint fermions \cite{Edwa99a}, which is in the class $\beta =1$, and
for strong coupling $U(1)$ gauge theory \cite{Bergu1} (also with Dyson
index $\beta =2$).
Although most results have been obtained in the quenched limit, the
agreement with chRMT for dynamical quark masses of order $1/V\Sigma$ 
\cite{Jurk96,poulmass,Wilk98,tilomass,gernot,Nishim}
or massless quarks in the Schwinger model \cite{Farc,Farc99a} 
is equally impressive.
\begin{figure}[!ht]
  \begin{center}
   \epsfig{figure=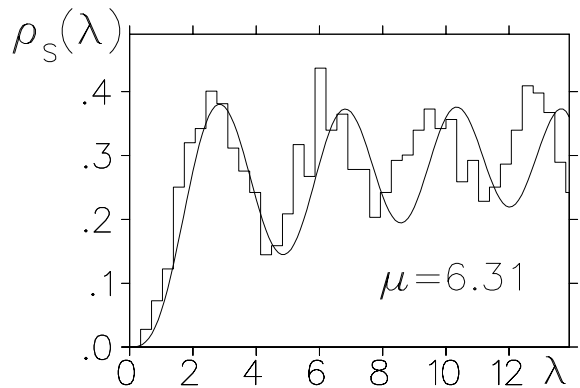,width=50mm,angle=90}
    \caption[]{Microscopic spectral density for nonzero dynamical quark mass
      for the staggered Dirac
      operator in $SU(2)$ \cite{tilomass}.  The dashed curve is the prediction
      \cite{gernot,Nishim} of the chSE for $\nu=0$. (From Ref. \cite{gernot}.)}
    \label{gernotfig}
  \end{center}
\end{figure}
In Fig. \ref{gernotfig} we show the microscopic spectral density
for dimensionless dynamical quark mass $\mu =m V \Sigma$ as given in the 
legend of the figure. 

\begin{figure}[!ht]
  \begin{center}
    \epsfig{figure=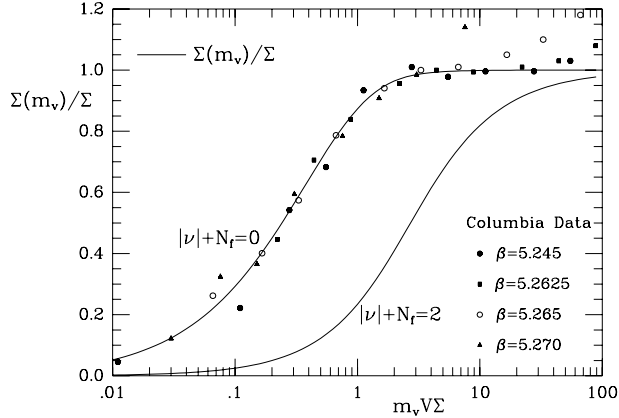,width=55mm,angle=270}
    \caption{Valence quark mass dependence of the chiral condensate
      $\Sigma(m_v)$ plotted as $\Sigma(m_v)/\Sigma$ versus $m_v
      V\Sigma$.  The dots and squares represent lattice results by the
      Columbia group 
\protect\cite{Chan95} 
    for the values of $\beta$ indicated
      in the figure.  The solid curves are chRMT results. 
(From Ref. \cite{Vplb}.)}
    \label{valence}
  \end{center}
\end{figure}

A second way to study the Dirac spectrum is by means of the valence
quark mass dependence of the chiral condensate defined by
\be
\Sigma(z) = \frac 1V \left \langle{\rm Tr} \frac 1{D+z} \right \rangle.
\ee
Its analytical expression 
can be derived from the low-energy limit of the QCD partition function
\cite{OTV,DOTV} as
well as from chRMT \cite{Vplb}. For $\beta =2 $ we find  \cite{Vplb}
\begin{equation}
  \label{valk}
  \frac{\Sigma(z)}{\Sigma} = x\left[I_{a}(x)K_{a}(x) 
    +I_{a+1}(x)K_{a-1}(x)\right]\:,
\end{equation}
where $ a= N_f+|\nu|$ and $x= zV\Sigma$. In Fig. \ref{valence} we compare
\cite{Vplb} the
analytical result (full line) with lattice data obtained by the
Columbia group \cite{Chan95}. The point above which the lattice data depart
from the the chRMT result agrees with our estimate of the Thouless
energy (\ref{thouless}). 
These results have been confirmed by independent simulations
\cite{Damval,Hernandezmv} and have been extended to other symmetry
classes \cite{Damval}.

\begin{figure}[!ht]
  \begin{center}
    \epsfig{figure=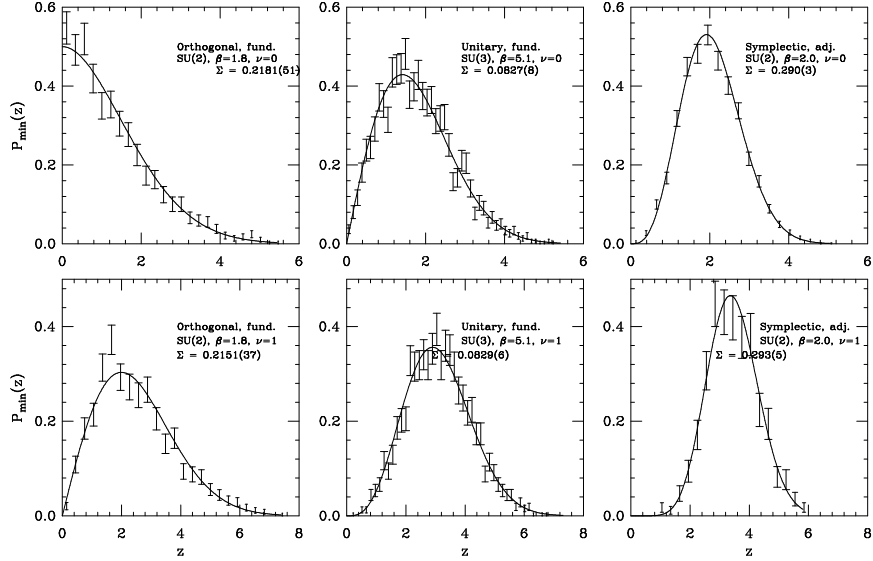,width=130mm}
\vspace*{-3cm}
    \caption{Distribution of the smallest Dirac eigenvalue for $\nu = 0$ and
      $\nu=1$  obtained from an  overlap Dirac operator on a $4^4$ lattice.
      The solid lines represent the chRMT results.  
      (From Ref.~\cite{hellertop}.)}
    \label{hellerfig}
  \end{center}
\end{figure}

A third way to analyze the statistical properties of the Dirac
eigenvalues is by means of the distribution of the smallest
eigenvalues. 
As an example, the
analytical result for $\beta =2 $  is given by \cite{Forr93},
$P(\lambda_{\rm min}) = \frac {\lambda_{\rm min}}2 
\exp({-\lambda_{\rm min}^2/4})$.
The results for $\beta =1 $, $\beta =4$, and nonzero
quark masses, are more complicated
\cite{Forr93,poulmass,tilomass,Nish98}. 
In Fig. \ref{figmicro} 
we show results \cite{Berbenni} for quenched $SU(2)$ lattice data. Results
for all three symmetry classes as well as nonzero topological
charge \cite{hellertop} are shown in Fig. \ref{hellerfig}. 
The latter results were
obtained with the overlap Dirac operator. Dirac spectra of the
Schwinger model have also been analyzed at nonzero topological charge
\cite{Farc} and complete agreement with chRMT was found \cite{Schnabel}.
The continuum limit of the staggered Dirac operators is approached
very slowly, and on today's lattices the Dirac spectra
are described by analytical results for zero topological charge \cite{Damg99c}.
Recently, analytical results  for the $k$'th smallest
eigenvalue \cite{nishi-k} gave a perfect description of the lattice data
\cite{poul-k}. 

A more subtle way to study the statistical properties of eigenvalues
is by means of the two-point correlation function defined as
\be
\rho(\lambda,\lambda') = \langle \sum_{k,l} \delta(\lambda - \lambda_k)
\delta(\lambda' - \lambda_l )\rangle.
\ee
The two-point correlation
function for the quenched $SU(2)$ staggered Dirac operator was
compared with chRMT in \cite{Berbenni,Ma98}. Also for instanton
liquid gauge field configuration one finds \cite{james} agreement with
chRMT in its domain of validity. The volume dependence of the
Thouless energy was investigated both for instanton liquid configurations
\cite{james,antonio} and lattice QCD simulations \cite{Berb98c,tilo3} and 
good agreement with the theoretical predictions
\cite{Vplb} was found.
A related quantity is the 
disconnected scalar susceptibility defined by
  \begin{equation}
  \chi^{\mathrm {disc}}(m)
    =\frac{1}{N}\left\langle\sum_{k,l=1}^N \frac{1}
      {(i\lambda_k+m)(i\lambda_l+m)}\right\rangle
     -\frac{1}{N} \left\langle\sum_{k=1}^N\frac{1}
      {i\lambda_k+m}\right\rangle^2 \:.
\end{equation}

\begin{figure}[!ht]
  \begin{center}
    \epsfig{figure=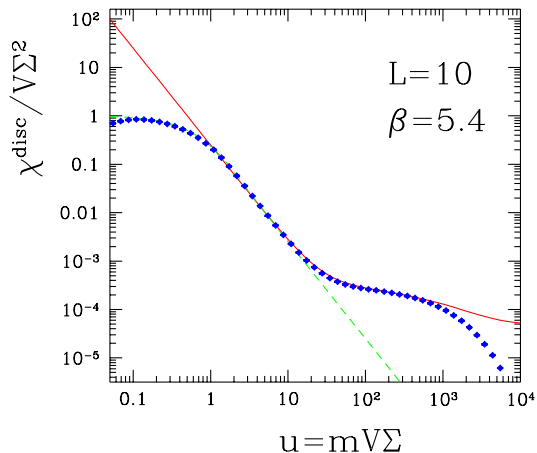,width=70mm}
    \caption{Comparison of the disconnected susceptibility 
      computed on the lattice in quenched SU(3) with staggered
      fermions ($V=10^4$, $\beta=5.4$) (points) with the
      prediction of chPT (full line) and the prediction of chRMT
    (dashed curve).
      (Note the dashed line is hidden by the data points for $u<10$.)
       (From Ref. \cite{Berb99}.)      }
    \label{chPT}
  \end{center}
\end{figure}

Lattice results \cite{Berb99} (see Fig. \ref{chPT}) show a sharp transition
point which can be identified as the Thouless energy. Below this
energy the susceptibility follows the 
chRMT prediction (dashed curve).
A complete analytical description up to $\Lambda_{\rm QCD}$ is
obtained from chiral perturbation theory (full curve) 
\cite{Berb99,kim2,dom2}
which applies in the domain  $\lambda_{\rm min} \ll m \ll \Lambda_{\rm QCD}$.
Similar agreement has been found for the scalar susceptibility of
the $U(1)$ staggered Dirac operator \cite{Bergu1}.

Dirac spectra at finite temperature and nonzero chemical potential
have been studied in much less detail. Because of finite size effects
comparisons with chRMT are difficult at the critical temperature. 
Dirac spectra
near $T_c$ were analyzed in detail in \cite{Damgaard:2000cx}. Because 
the chiral phase transition in chRMT has mean field critical exponents
\cite{JV,tilot} there is no reason to believe that the eigenvalue
fluctuations follow the Random Matrix predictions \cite{Akem98,hikami,Jani98}
at the critical point. 
However, beyond $T_c$
there is some evidence that  the smallest eigenvalues show 
a fluctuation behavior as predicted by RMT \cite{Farc99b}. 
At nonzero chemical potential the Dirac operator is nonhermitian and
its eigenvalues are scattered in the complex plane. Recent
work \cite{hands,azco} shows that the global
spectral properties are described by a chiral Lagrangian
\cite{kstvz,many} or Random Matrix Theory \cite{Misha}.
The statistical
analysis of the eigenvalue fluctuations is much more complicated in this case,
but the first lattice results \cite{Tilomu} seem to confirm the
theoretical expectations 
\cite{fyodorovpoly,fyodorov,yanzeros,Efetovnh,efetovrot}.

\begin{figure}[!ht]
  \begin{center}
    \epsfig{figure=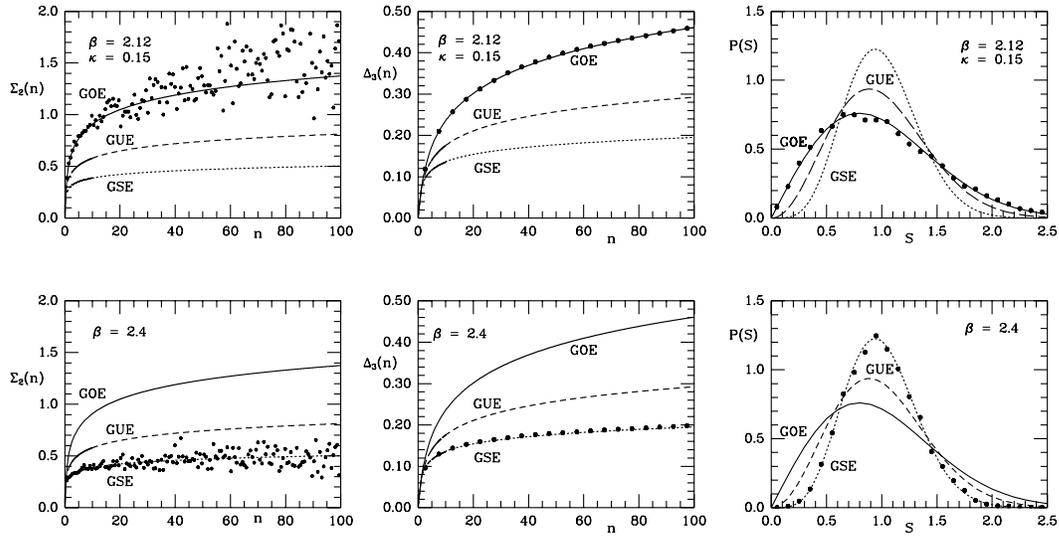,angle=270,width=\textwidth}
    \caption{Number variance $\Sigma^2(n)$, the 
      $\Delta_3(n)$ statistic, 
      and nearest-neighbor spacing distribution $P(S)$
      of the $SU(2)$ lattice Dirac operator.  Upper
      row: Wilson fermions, $V=8^3\times12$, $N_f=2$.  Lower row:
      staggered fermions, $V=12^4$, $N_f=4$, $ma=0.05$.  (From
      Ref.~\protect\cite{HKV}.)} 
    \label{bulk}
  \end{center}
\end{figure}
\begin{figure*}[!ht]
  \centerline{\epsfig{figure=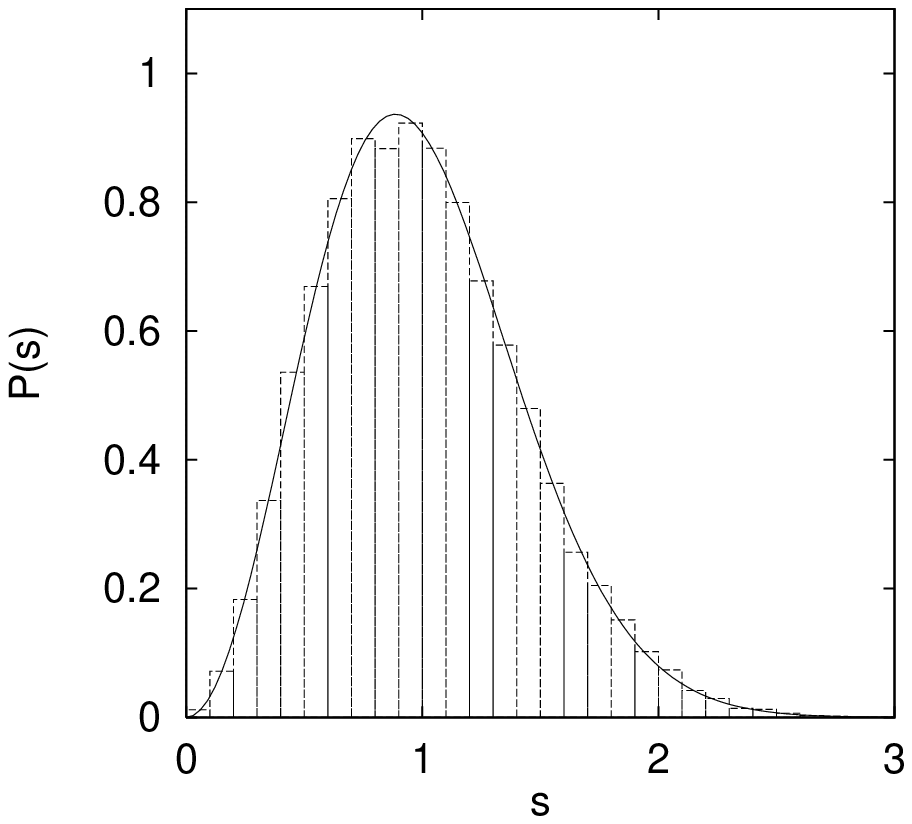,width=4.4cm}\hspace*{3mm}
    \epsfig{figure=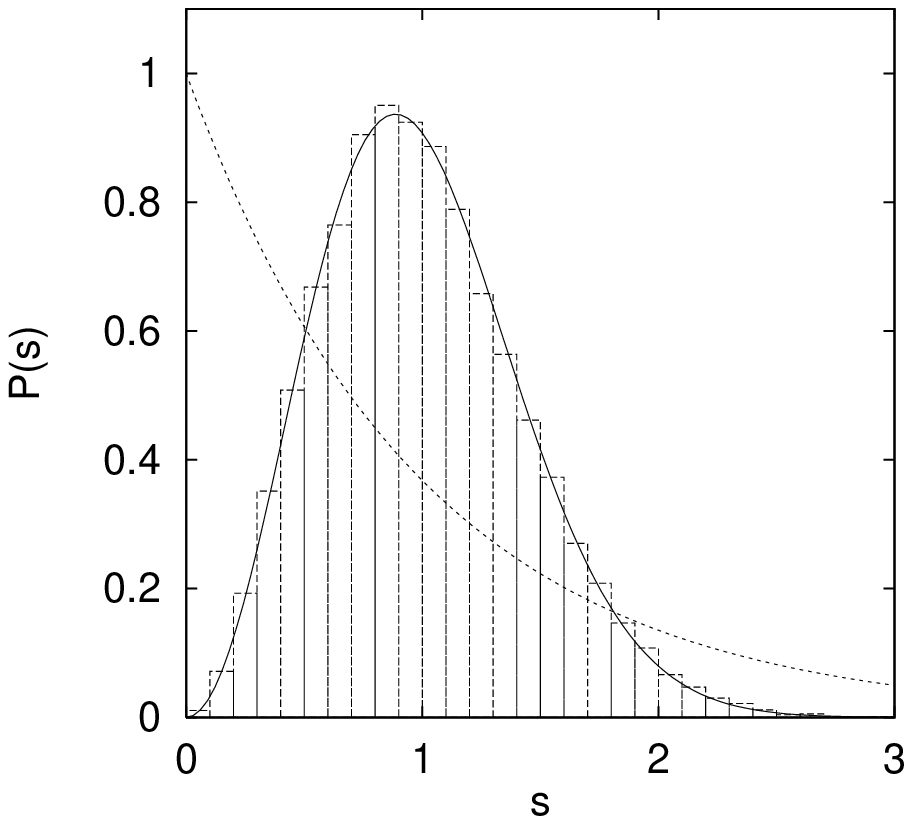,width=4.4cm}\hspace*{3mm}
    \epsfig{figure=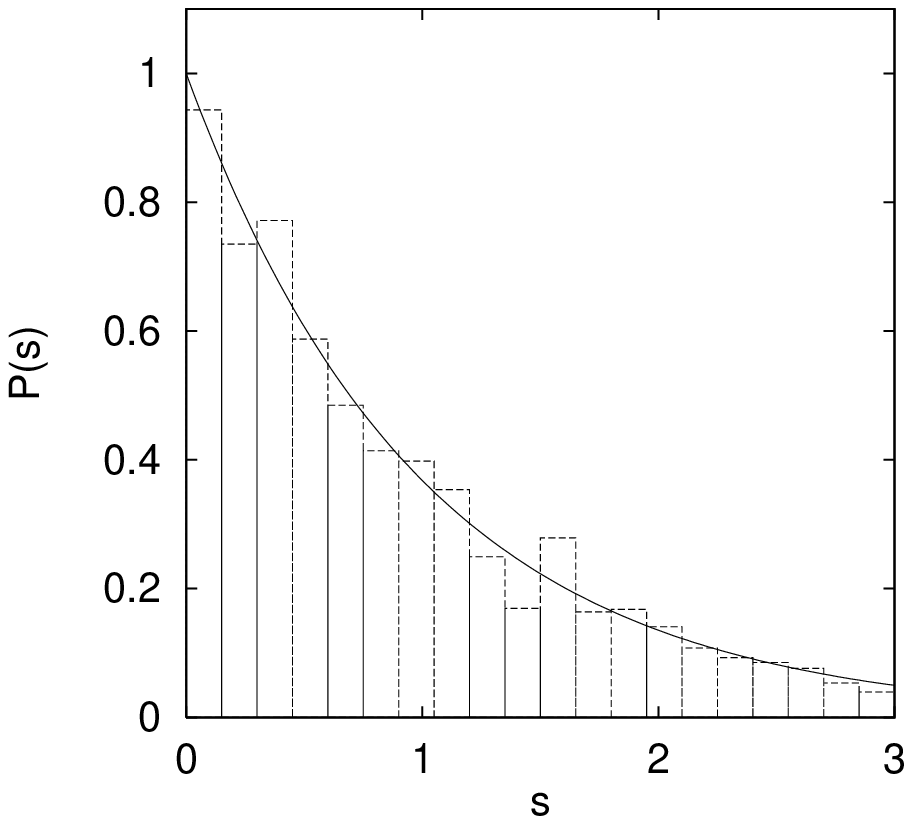,width=4.4cm}}
  \caption{Nearest neighbor spacing distribution $P(s)$ on  an
    $8^3\times 6$ lattice in the confined phase (left) in the
    deconfined phase (middle), and for the free Dirac operator on a
    $53\times 47\times 43\times 41$ lattice (right). The curves represent
    the analytical chGUE result,
    $P(s)=32(s/\pi)^2\exp(-4s^2/\pi)$, and the Poisson distribution,
    $P(s)=\exp(-s)$. (From Ref. \cite{pisa}.)}  
  \label{trans}
\end{figure*}

Up to now we only discussed the statistical properties of the Dirac
eigenvalues near $\lambda =0$. Although physically less relevant, one
can also analyze the statistics of the eigenvalues in the bulk of the
spectrum. Assuming that the statistical properties do not change 
along the spectrum we can average in two different ways: by averaging over
independent gauge field configurations and by averaging over the
spectrum. The advantage of spectral averaging is that it requires only
one or a few independent gauge field configurations.
The equality of the two averages is known as spectral 
ergodicity and was investigated in the context of QCD Dirac spectra in
\cite{Guhr99}. The Thouless energy was only found
in ensemble averaging. The enhanced eigenvalue fluctuations 
result from the ``collective'' motion of the eigenvalues in the
evolution of the ensemble. In Fig. \ref{bulk} we show \cite{HKV}
the spacing distribution $P(S)$ of neighboring eigenvalues,
the number
variance $\Sigma^2(n)$ and the $\Delta_3$ statistic. The number
variance is defined as the variance of the number of levels in an
interval containing $n$ eigenvalues on average, and  $\Delta_3(n)$
is obtained by integrating $\Sigma^2(n)$  over a 
smoothening kernel. The question has been raised whether eigenvalue
correlations in the bulk are different above and below  $T_c$, but
no effects have been seen \cite{Pull98,Berg99,Bergu1,pisa} 
(see also Fig. \ref{trans}). 
A transition toward Poisson statistics only takes place at very small
values of the coupling constant (see Fig. \ref{trans}). 

Finally, lattice results for QCD Dirac operator in 3 dimensions
\cite{Damg98b} have been compared with Random Matrix Theory. In this
case one finds agreement with the Wigner-Dyson ensembles
\cite{Damg98b},
but we will not discuss this topic in this review.

\section{Conclusions}
The generating function of the Dirac spectrum is given by the QCD
partition function with additional ghost quarks with a mass scale given by
the region of the the Dirac spectrum we are interested in. 
At low energies this partition function reduces to a gas of
weakly interacting Goldstone modes. In the domain where the kinetic
term can be neglected it reduces to  a chiral Random Matrix Theory 
with the global symmetries of the QCD partition function. The predictions
based on these arguments have been confirmed by numerous lattice QCD
simulations. 
This does not mean that one can refrain from doing lattice simulations.
The point is that universal behavior and the phenomenologically  relevant
nonuniversal properties are found in the same lattice simulations. The real
progress is the understanding of the symbiosis of these two features of 
the strong interactions.

\end{document}